\documentclass[10pt,aps,prc,twocolumn,floatfix,showpacs]{revtex4-1} 
\usepackage{graphics}
\usepackage{graphicx}
\usepackage{sidecap}
\usepackage{array}
\usepackage{subfigure}
\usepackage{color}
\usepackage{lineno}
\usepackage{placeins}
\usepackage{cleveref}

\usepackage{ulem}

\begin{document}%

\title{Blast Wave Fits to Elliptic Flow Data at $\sqrt{s_{\rm NN}} =$ 7.7--2760 GeV}

\author{X.~Sun$^{1,2}$}
\author{H.~Masui$^{3}$}
\author{A.~M.~Poskanzer$^{2}$}
\author{A.~Schmah$^{2}$}

\address{$^{1}$Department of Physics, Harbin Institute of Technology, Harbin 150001, People's Republic of China}
\address{$^{2}$Nuclear Science Division, Lawrence Berkeley National Laboratory, Berkeley, California 94720, USA}
\address{$^{3}$Institute of Physics, University of Tsukuba, Tsukuba, Ibaraki 305, Japan}

\date{\today}

\realpagewiselinenumbers
\setlength\linenumbersep{0.10cm}

\begin{abstract}
We present blast wave fits to elliptic flow ($v_{2}(p_{\rm T})$) data in minimum bias collisions from $\sqrt{s_{\rm NN}} =$ 7.7--200 GeV at RHIC, and also at the LHC energy of 2.76 TeV. The fits are performed separately for particles and corresponding anti-particles. The mean transverse velocity parameter $\beta$ shows an energy dependent difference between particles and corresponding anti-particles, which increases as the beam energy decreases. Possible effects of feed down, baryon stopping, anti-particle absorption, and early production times for anti-particles are discussed.
\end{abstract}

\maketitle
%

\section{Introduction} 
\label{sec_intro}

To understand the formation of the Quark-Gluon Plasma (QGP) phase and to study the structure of the QCD phase diagram, a Beam Energy Scan (BES) program has been carried out in the years 2010 and 2011 at the Relativistic Heavy Ion Collider (RHIC) facility~\cite{Aggarwal:2010cw} where Au+Au collisions were recorded at $\sqrt{s_{\rm NN}} =$  7.7, 11.5, 19.6, 27, 39, and 62.4 GeV. Azimuthal anisotropy~\cite{Voloshin:2008dg} is one of the most important observables in relativistic nuclear collisions for studying the bulk behavior of the created matter. In non-central Au+Au collision, the overlap region has an almond shape (in each event) with the major axis perpendicular to the reaction plane, which is defined by the impact parameter and the beam direction. Due to fluctuations, the participant plane~\cite{PartPlane} in each event is not necessarily the same as the reaction plane. As the system evolves, the pressure gradient converts the anisotropy from coordinate space to momentum space. The produced particle distribution~\cite{Voloshin:1994mz,Poskanzer:1998yz} can be written as:

\begin{equation}
\label{eq:EP}
E\frac{d^{3}N}{dp^{3}} = \frac{1}{2\pi}\frac{d^{2}N}{p_{\rm T}dp_{\rm T}dy}(1+\sum_{n=1}^{\infty}2v^{\rm obs}_{n}\cos[n(\phi-\Psi_{n})]),
\end{equation}

\begin{equation}
\label{eq:res}
v_{n} = v_{n}^{\rm{obs}}/R_{n},
\end{equation}
where $\phi$ is the azimuthal angle of a particle, $\Psi_{n}$ is the $n$-th harmonic event plane angle reconstructed by the observed particles, which is an estimation of the participant plane, and $R_{n}$ is the $n$-th harmonic event plane resolution.
The second harmonic coefficient $v_{2}$ reported here is called elliptic flow.

Several interesting observations related to $v_{2}$ have been reported in the past decade by using data from the top RHIC heavy-ion collisions energy of $\sqrt{s_{\rm NN}} =$ 200 GeV~\cite{Voloshin:2008dg,Borghini:2004ra,Sorensen:2009cz,Adams:2005dq,Adcox:2004mh}. At low transverse momenta ($p_{\rm T} < $ 2.0 GeV/$c$), a mass ordering of the $v_{2}$ values was observed~\cite{Adler:2001nb,Adams:2003am,Adams:2005zg}, which could be understood within a hydrodynamic framework. At intermediate $p_{\rm T}$, (2 $< p_{\rm T} <$ 6 GeV/$c$), a Number-of-Constituent Quark (NCQ) scaling~\cite{NCQScaling} of $v_{2}$ for identified hadrons was observed. This observation, coupled with comparable values of the elliptic flow measured for multi-strange hadrons ($\phi$ and $\Xi$) and light quark hadrons, was used to conclude that the relevant degrees of freedom are quarks and gluons for the matter formed in the early stage of heavy ion collisions at the top RHIC energy~\cite{Abelev:2007rw,Abelev:2010tr,Adams:2005zg,Abelev:2007qg,Voloshin:2008dg}.

The mass ordering in the low $p_{\rm T}$ range and the NCQ scaling in the intermediate $p_{\rm T}$ range were also observed in BES experiments~\cite{Adamczyk:prc}. In this paper we use the blast wave model~\cite{Westfall:1976,Siemens:1979, Schnedermann:1993, Huovinen:2001, STAR:2001bw} to fit $v_{2}(p_{\rm T})$ data at $\sqrt{s_{\rm NN}} =$ 7.7 -- 2760 GeV to get the energy dependence of the mean radial flow expansion velocity. The blast wave model is an approximation to the full hydro calculations, which were only done for BES inclusive charged hadron data~\cite{HydroV2}, not for identified particles due to complications of the equation-of-state and the initial conditions. 

This paper is organized as follows. Section~\ref{sec_blast} gives a brief introduction to the blast wave model and the fit functions used in this paper. In Section~\ref{sec_results} we show the fit results and discuss the physics implications. A summary is given in Section~\ref{sec_summary}.

\section{Blast Wave Parametrization} 
\label{sec_blast}

The nuclear fireball model~\cite{Westfall:1976} was introduced by Westfall {\it et al.} to explain mid-rapidity proton inclusive spectra. This model assumes that a clean cylindrical cut is made by the projectile and target and leaves a hot source in between them. Protons emitted from this fireball should follow a thermal energy distribution.  
Later, Siemens and Rasmussen~\cite{Siemens:1979} generalized a formula by Bondorf, Garpman, and Zimanyi~\cite{Bondorf:1978} which was valid for non-relativistic velocities, to be fully relativistic assuming an exploding fireball producing a blast wave of nucleons and pions.  
Two decades ago, Schnedermann {\it et al.}~\cite{Schnedermann:1993} introduced a simple functional form with only two fit parameters: a kinetic temperature ($T$) and a radial velocity ($\beta$) which was successfully used in fits to $p_{\rm T}$ spectra.  
Huovinen {\it et al.}~\cite{Huovinen:2001} introduced a third parameter, the difference of the radial velocity in and out of the reaction plane, to describe transverse anisotropic flow generated in non-central collisions.

However, the blast wave fit matched data even better after the STAR Collaboration added a fourth parameter~\cite{STAR:2001bw} to take into account the anisotropic shape of the source in coordinate space.

We use the blast wave parametrization with four parameters mentioned above~\cite{STAR:2001bw}: kinetic freeze-out temperature ($T$), transverse expansion rapidity ($\rho_{0}$), the amplitude of its azimuthal variation ($\rho_{a}$) and the variation in the azimuthal density of the source elements ($s_{2}$).
The blast wave equation we use is:

\begin{widetext}
\begin{equation}
\label{eq:v2}
v_{2}(p_{t}) = \frac{\int_{0}^{2\pi}d\phi_{s} \cos(2\phi_{s})I_{2}[\alpha_{t}(\phi_{s})]K_{1}[\beta_{t}(\phi_{s})][1+2s_{2}\cos(2\phi_{s})]}{\int_{0}^{2\pi}d\phi_{s} I_{0}[\alpha_{t}(\phi_{s})] K_{1}[\beta_{t}(\phi_{s})][1+2s_{2}\cos(2\phi_{s})]}.
\label{lab_BW_func}
\end{equation}
\end{widetext}
$I_{0},\ I_{2}$, and $K_{1}$ are modified Bessel functions where $\alpha_{t}(\phi_{s}) = (p_{\rm T}/T) \sinh[\rho(\phi_{s})]$, and $\beta_{t}(\phi_{s}) = (m_{T}/T)\cosh[\rho(\phi_{s})]$. It should be noticed that the masses for different particle species only enter via $m_{T}$ in $\beta_{t}(\phi_{s})$. When we perform the simultaneous fits, which will be explained below, the masses for different particle species are the only differences between the fits to different particle species. The basic assumptions of this blast wave model is a boost-invariant longitudinal expansion~\cite{Bjorken:1983} and freeze-out at a constant temperature $T$ on a shell~\cite{cooper_frye}, which expands with transverse rapidity exhibiting a second harmonic azimuthal modulation given by $\rho(\phi_{s}) = \rho_{0} + \rho_{a}\cos2\phi_{s}$~\cite{STAR:2001bw}. In this equation, $\phi_{s}$ is the azimuthal angle in coordinate space and $\beta = \tanh(\rho_{0})$, where $\beta$ is the transverse expansion velocity.

\section{Results} 
\label{sec_results}
We present simultaneous blast wave fit results for $v_{2}(p_{\rm{T}})$ for a particle group ($K^{+}, K^{0}_{s}, p$, $\phi$ and $\Lambda$) and for an anti-particle group ($K^{-}, K^{0}_{s}, \bar{p}$, $\phi$ and $\bar{\Lambda}$) from 0\%--80\% central Au+Au collisions at $\sqrt{s_{\rm{NN}}} =$ 7.7--200 GeV~\cite{Adamczyk:prc}. The data at $\sqrt{s_{\rm{NN}}} =$ 2.76 TeV~\cite{ALICE:PIDv2} covers only 0\%--60\% in centrality and was merged from finer centrality selections using particle spectra from~\cite{ALICE:PiKP,ALICE:Lambda}. At $\sqrt{s_{\rm{NN}}} =$ 200 GeV and 2.76 TeV particles and anti-particles were merged due to their small difference in $v_{2}(p_{\rm T})$.

\subsection{Fit Procedure}
\label{sec_procedure}

\begin{figure}[]
\centering
\resizebox{8cm}{!}{%
\includegraphics[scale=1]{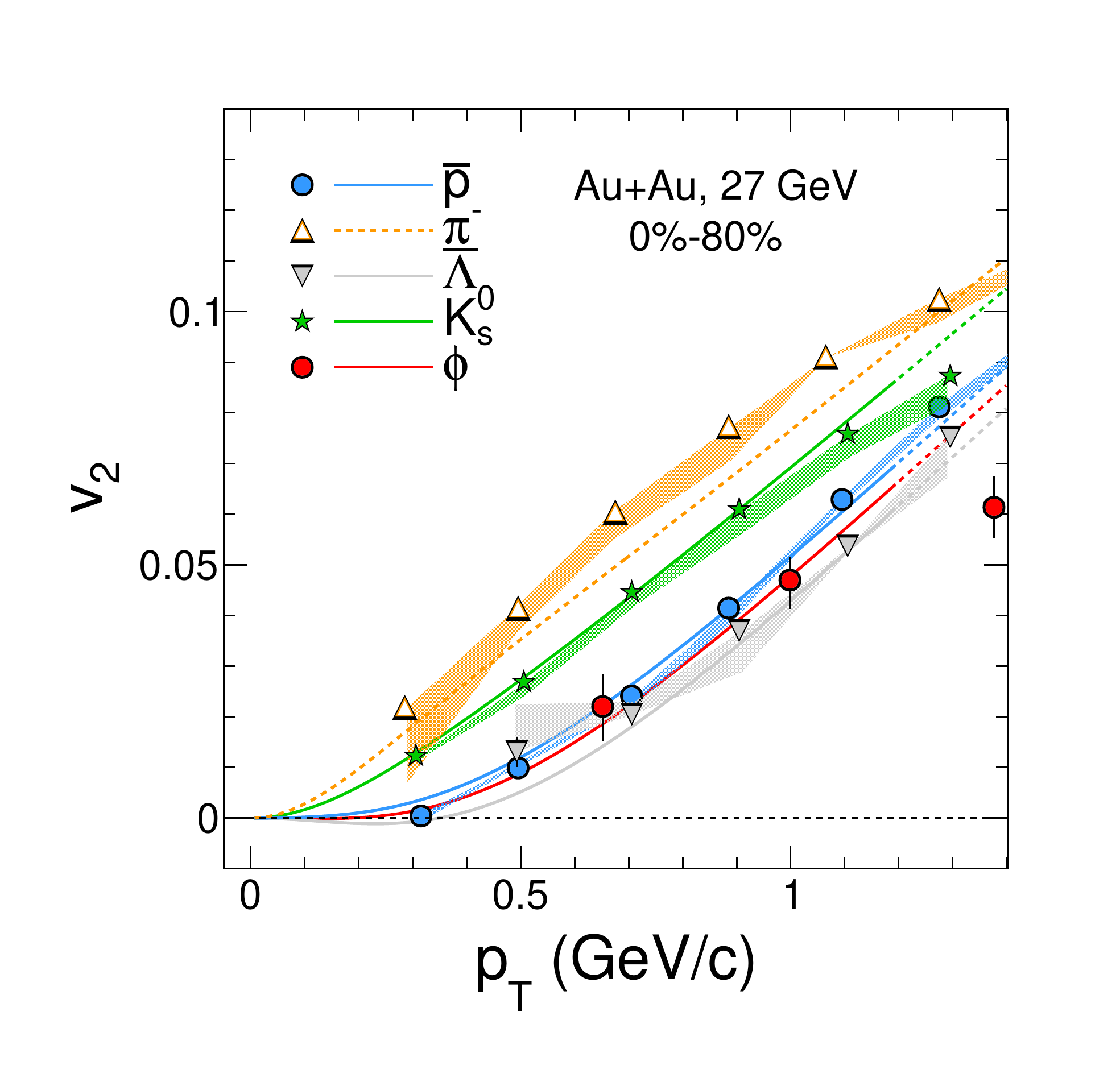}}
\caption{\label{fig:feeddown} (Color online) Elliptic flow ($v_{2}$) as a function of transverse momentum $p_{\rm T}$ for Au+Au collisions at $\sqrt{s_{\rm{NN}}} =$ 27 GeV for a selected group of particles. The shaded areas show estimations for the feed down correction. Solid lines are from blast wave fits and dashed lines are predictions by using the fit parameters.} 
\end{figure}

\begin{figure*}[ht]
\begin{minipage}[b]{0.48\linewidth}
\centering
\makebox[\linewidth]{%
\includegraphics[width=1.0\textwidth]{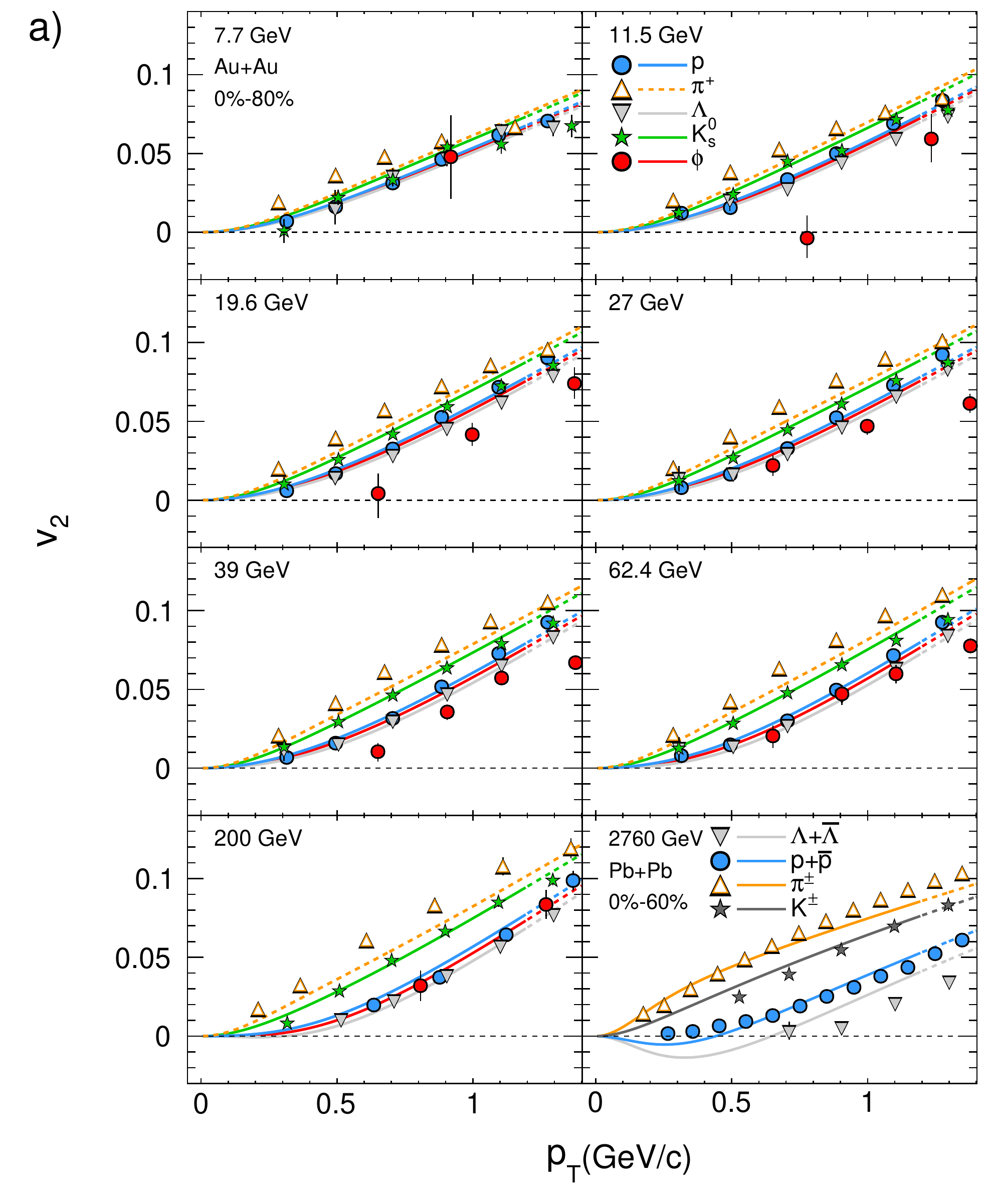}}
\end{minipage}
\hspace{0.5cm}
\begin{minipage}[b]{0.48\linewidth}
\centering
\makebox[\linewidth]{%
\includegraphics[width=1.0\textwidth]{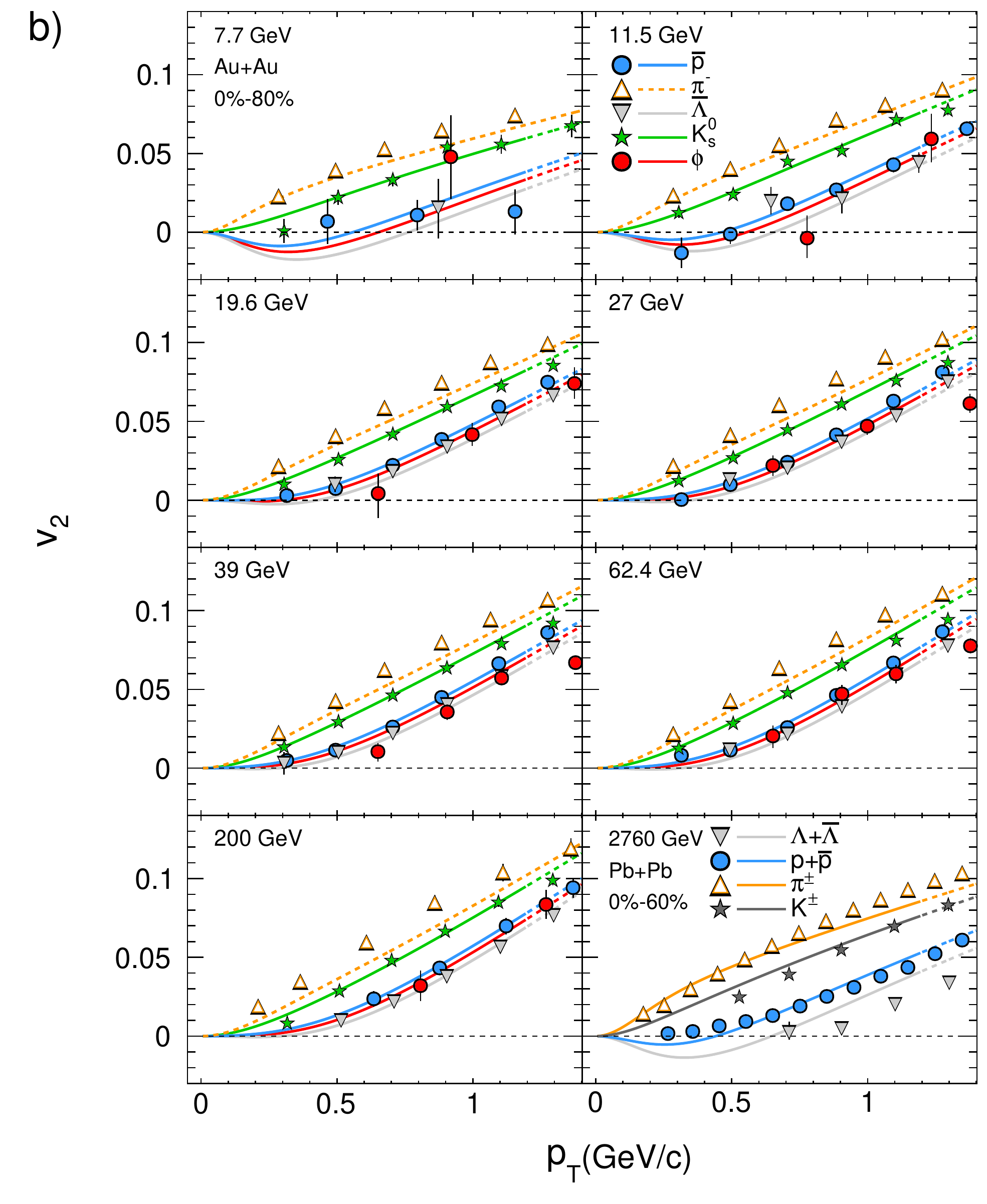}}
\end{minipage}
\caption{\label{fig:fit} (Color online) The simultaneous blast wave fits for $v_{2}$ of a) particles ($K^{0}_{s}, p$ and $\Lambda$) and b) corresponding anti-particles ($K^{0}_{s}, \bar{p}$ and $\bar{\Lambda}$) from 0\%--80\% central Au+Au collisions at $\sqrt{s_{\rm{NN}}} =$ 7.7--200 GeV and for combined $v_{2}$ of particles and anti-particles ($\pi^{\pm}, K^{\pm}, p+\bar{p}$ and $\Lambda+\bar{\Lambda}$) from 0\%--60\% central Pb+Pb collisions at $\sqrt{s_{\rm{NN}}} =$ 2.76 TeV. Solid lines are from blast wave fits and dashed lines are predictions by using the fit parameters.}
\end{figure*}

We fit $v_{2}(p_{\rm T})$ data from 0\%--80\% (0\%--60\% for ALICE data) central Au+Au (Pb+Pb) collisions with Eq.~(\ref{lab_BW_func}). Fits are done only for $p_{\rm T} < 1.2\ {\rm GeV}/c$ to avoid the jet contributions at high $p_{\rm T}$. Furthermore the fits are separated for the particle group ($K^{+}, K^{0}_{s}, p, \phi$ and $\Lambda$) and the anti-particle group ($K^{-}, K^{0}_{s}, \bar{p}, \phi$ and $\bar{\Lambda}$), as we know from the BES experiments that their $v_{2}$ values at the same $p_{\rm T}$ are different~\cite{Adamczyk:2013}. $K^{0}_{s}$ and $\phi$ mesons are used twice for both particles and anti-particles.

All $v_{2}(p_{\rm T})$ data in each group are fitted simultaneously. For the fits, statistical and systematic errors of the data were added in quadrature. As we do not have spectra for most of the energies, we cannot constrain the temperature; therefore we input a temperature in a reasonable range~\cite{Tkin}. In this paper we choose $T =$ 100 MeV, 120 MeV and 140 MeV as input. The fit lines in the following figures are for $T =$ 120 MeV, the other two temperatures are considered to determine the systematic variation and are shown in the summary Fig.~\ref{fig:beta}.

\subsection{Feed Down}
\label{sec_feeddown}

Pions are excluded from the fits at energies below $\sqrt{s_{\rm{NN}}} =$ 2.76 TeV, as a significant fraction of the pions at those energies come from resonance decays~\cite{resonance:pion}, and therefore might behave differently from the expected blast wave parametrization. The ALICE collaboration has reduced the feed-down contributions from long lived particles to their $\sqrt{s_{\rm NN}}$ = 2.76 TeV results by selecting tracks with a small distance of closest approach to the primary event vertex~\cite{ALICE:PiKP}. 

Other particles, like protons and kaons, are also affected by feed down from heavier particles and would therefore show deviation from the blast wave parametrization. For a correct feed down correction one needs particle spectra, an estimation of resonance production, and a detailed understanding of the kinematic and topological cuts applied. Since this information is not yet available, we therefore estimated the feed down effects with a Monte Carlo (MC) calculation. The inputs for the MC were particle and resonance yields estimated from the THERMUS statistical hadronization model~\cite{THERMUS}, and we used a parametrization~\cite{StatPar} for the energy dependence of the chemical freeze-out temperature $T$ and baryon chemical potential $\mu_{B}$. We further used Boltzmann distributions to sample the transverse momenta of the parent particles. The flow of resonances was estimated using the blast wave fits. This implies that an iterative process would be needed for a more detailed study. The decay kinematics of various particles and resonances ($\phi$, $\Lambda$, $\Xi$, $\Omega$, $\Delta(1232)^{++}$, $\omega$, N$^{*}$, $\Sigma^{0}$) was calculated to get the modified flow and transverse momenta of the daughter particles which contribute to the measured $v_{2}(p_{\rm T})$. For pions a total feed down contribution from resonances of 60\% was used~\cite{THERMUS}. 

An example for feed down corrected $v_{2}(p_{\rm T})$ is given in Fig.~\ref{fig:feeddown}. One can see that the feed down correction is significant for all particles and exceeds the statistical and systematic errors of the data. Therefore one cannot expect to get a perfect description of the data with the blast wave model. 

Figure~\ref{fig:feeddown} shows the size of the effect, but for the fits shown in this paper no feed down corrections were applied due to the uncertainties as discussed above. However, to estimate the feed-down effect on the fit results and the $\chi^{2}$/ndf we redid all fits by root-mean-square adding to every data point a $v_{2}$ value of 0.003 for $s_{\rm NN} = $ 7.7--200 GeV based on our feed-down studies. All resulting changes of the fit results turned out to be smaller than the shown statistical error bars. 

\begin{table*}[htc]  
  \caption{\label{table} Fit parameters $\rho_{\rm 0}$, $\rho_{a}$ and $s_{\rm 2}$ for the particle group ($\rm X$) and the anti-particle group ($\bar{\rm X}$) from min.-bias Au+Au collisions at $\sqrt{s_{\rm NN}} = $ 7.7 -- 200 GeV and Pb+Pb collisions at $\sqrt{s_{\rm NN}} = $ 2760 GeV.} 
  \footnotesize\rm  
  \begin{tabular}{ >{\centering\arraybackslash}m{0.7in}  >{\centering\arraybackslash}m{0.70in} >{\centering\arraybackslash}m{0.70in} >{\centering\arraybackslash}m{0.70in} >{\centering\arraybackslash}m{0.70in} >{\centering\arraybackslash}m{0.70in} >{\centering\arraybackslash}m{0.70in} >{\centering\arraybackslash}m{0.70in} >{\centering\arraybackslash}m{0.70in}}  
    \hline \hline  
    $ $&$ 7.7\ \rm GeV $&$ 11.5\ \rm GeV $&$ 19.6\ \rm GeV $&$ 27\ \rm GeV $&$ 39\ \rm GeV $&$ 62.4\ \rm GeV $&$ 200\ \rm GeV $&$ 2760\ \rm GeV $\\ 
    $\rho_{0} (\times 10^{-2} \ \rm X)$&$ 0.38 \pm 0.04 $&$ 0.42 \pm 0.01 $&$ 0.44 \pm 0.01 $&$ 0.47 \pm 0.01 $&$ 0.49 \pm 0.01 $&$ 0.51 \pm 0.01 $&$ 0.55 \pm 0.02 $&$ 0.89 \pm 0.02 $\\ 
    $\rho_{0} (\times 10^{-2} \ \bar{\rm X})$&$ 0.93 \pm 0.14 $&$ 0.77 \pm 0.04 $&$ 0.63 \pm 0.01 $&$ 0.59 \pm 0.01 $&$ 0.58 \pm 0.01 $&$ 0.57 \pm 0.01 $&$ 0.55 \pm 0.02 $&$ 0.89 \pm 0.02 $\\ 

    $ $&$  $&$  $&$  $&$ $&$ $&$ $&$ $&$ $\\ 
    $\rho_{a} (\times 10^{-2} \ \rm X)$&$ 2.73 \pm 0.28 $&$ 3.48 \pm 0.14 $&$ 3.79 \pm 0.07 $&$ 3.72 \pm 0.05 $&$ 4.03 \pm 0.03 $&$ 4.35 \pm 0.04 $&$ 4.62 \pm 0.29 $&$ 3.02 \pm 0.10 $\\ 
    $\rho_{a} (\times 10^{-2} \ \bar{\rm X})$&$ 2.56 \pm 0.37 $&$ 3.51 \pm 0.18 $&$ 3.75 \pm 0.08 $&$ 4.00 \pm 0.05 $&$ 4.11 \pm 0.03 $&$ 4.49 \pm 0.05 $&$ 4.66 \pm 0.29 $&$ 3.02 \pm 0.10 $\\ 

    $ $&$  $&$  $&$  $&$ $&$ $&$ $&$ $&$ $\\ 
    $s_{2} (\times 10^{-2} \ \rm X)$&$ 3.13 \pm 0.67 $&$ 2.36 \pm 0.31 $&$ 2.27 \pm 0.15 $&$ 2.74 \pm 0.10 $&$ 2.42 \pm 0.07 $&$ 2.17 \pm 0.09 $&$ 1.79 \pm 0.62 $&$ 4.62 \pm 0.11 $\\ 
    $s_{2} (\times 10^{-2} \ \bar{\rm X})$&$ 3.35 \pm 0.73 $&$ 3.17 \pm 0.32 $&$ 2.62 \pm 0.15 $&$ 2.35 \pm 0.10 $&$ 2.62 \pm 0.06 $&$ 2.17 \pm 0.09 $&$ 1.75 \pm 0.63 $&$ 4.62 \pm 0.11 $\\ 
    \hline  
  \end{tabular}  
\end{table*}  

Without feed-down correction the $\chi^{2}$/ndf of the fits is only close to 1 at lower energies, where the statistical errors are the order of the expected feed down effect. At higher energies the error bars are much smaller, the resulting $\chi^{2}$/ndf rises up to a maximum of 35 for the particle group at $\sqrt{s_{\rm{NN}}} =$ 39 GeV, whereas it is below 1.5 for all energies when feed-down contributions are included into the error bars. In the following we quote $\chi^{2}$/ndf values for the fits. The  $\chi^{2}$/ndf values with estimated feed-down contributions taken into account are shown in parenthesis. For anti-particles the $\chi^{2}$/ndf is systematically lower compared to the particle group with a maximum of 17 (1.5) at $\sqrt{s_{\rm{NN}}} =$ 39 GeV. At $\sqrt{s_{\rm{NN}}} =$ 200 GeV the $\chi^{2}$/ndf is again below 2 (0.4) due to large statistical error bars. At $\sqrt{s_{\rm{NN}}} =$ 2.76 TeV we have a description with a $\chi^{2}$/ndf of 9 (1.5). 

\subsection{Fit Results}
\label{sec_fit}

\begin{figure*}[ht]
\begin{minipage}[b]{0.45\linewidth}
\centering
\makebox[\linewidth]{%
\includegraphics[width=1.04\textwidth]{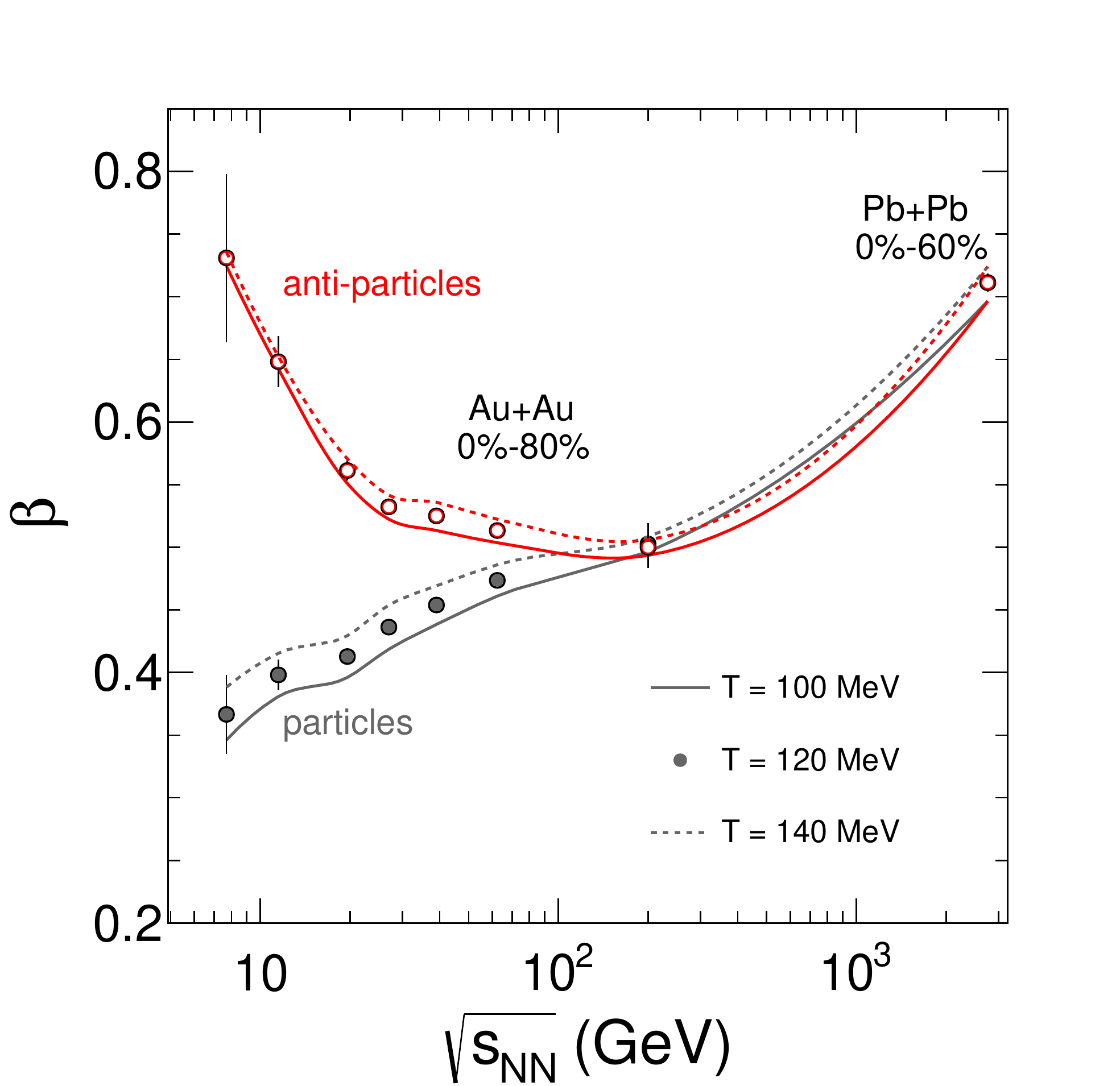}}
 \caption{\label{fig:beta} (Color online) The transverse expansion velocity $\beta$ as a function of beam energy from 0\%--80\% central Au+Au collisions and 0\%--60\% central Pb+Pb collisions for particles and anti-particles with three different temperatures.}
\end{minipage}
\hspace{0.5cm}
\begin{minipage}[b]{0.45\linewidth}
\centering
\makebox[\linewidth]{%
\includegraphics[width=1.04\textwidth]{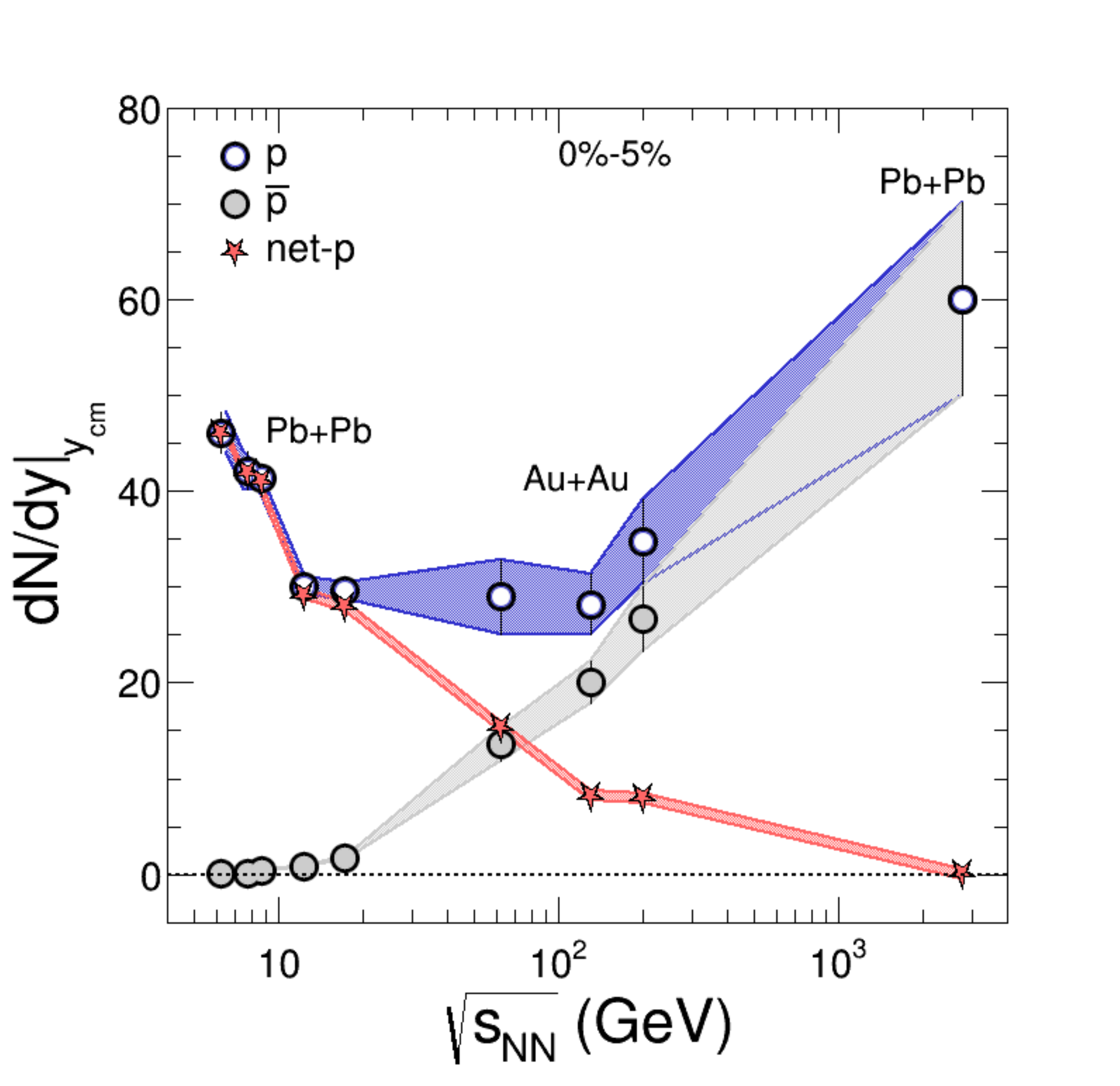}}
  \caption{\label{fig:netp} (Color online) Integrated multiplicity rapidity density of protons, anti-protons and net-protons as a function of the center-of-mass energy. Data is taken from Refs.~\cite{ALICE:PiKP,Yield:STAR,Yield:NA49}.}
\end{minipage}
\end{figure*}

Figure~\ref{fig:fit} a) shows the simultaneous blast wave fits for $v_{2}(p_{\rm T})$ of particles ($K^{0}_{s}, K^{+}, p, \phi$ and $\Lambda$) from 0\%--80\% (0\%--60\%) central Au+Au (Pb+Pb) collisions at $\sqrt{s_{\rm NN}} =$ 7.7--2760 GeV.
Solid lines depict blast wave fits to the data, whereas dashed lines are predictions, for pions, using the parameters from the fits to the other particles. The data points and fit curves for charged kaons are not shown in the figures as they are similar to the $K^{0}_{s}$ mesons. A clear mass ordering in data and fits is observed for all energies: for the same radial flow, the heavier particles have larger $p_{\rm T}$ values and therefore, at the same $p_{\rm T}$ lighter particles have larger $v_{2}$ values. If we assume that this splitting is due to radial flow, then the boost in the $p_{\rm T}$ direction gets larger with increasing beam energy, which is equivalent to a larger radial flow.

In general a fair description for protons and kaons can be obtained. On the other hand we observe that for all energies the predicted curves for $\pi^{+}$ have similar trends as the data points but are systematically lower. It cannot be excluded that such a behavior for pions is a result from feed down as discussed in section~\ref{sec_feeddown}.

The fit curves for $\phi$-meson are higher than the data points for all energies except for $\sqrt{s_{\rm NN}} =$ 62.4 GeV. It is argued that $\phi$-mesons have a small hadronic cross sections~\cite{phi:xsec,phi:review} compared to other hadrons under consideration. In that case, one expects a lower $\phi$-meson $v_{2}$ and therefore also a deviation from the blast wave fits. We want to point out that the weight of the $\phi$-meson data in the simultaneous fits is low due to their relatively large error bars. A fit without $\phi$-mesons included gives almost identical results.

In Fig.~\ref{fig:fit} b) we show the corresponding results for the anti-particle group ($K^{0}_{s}, K^{-}, \bar{p}, \phi$ and $\bar{\Lambda}$). The $K^{-}$ data is not shown for the same reason as for the $K^{+}$ mesons. The data show a larger spread along the $v_{2}$ or $p_{\rm T}$ axes, respectively, compared to the particle group. The simultaneous fits to all anti-particles are significantly better. Even trends like the negative values for anti-protons in the low $p_{\rm T}$ range at $\sqrt{s_{\rm NN}} = $ 11.5 GeV are reproduced. Similar to the particle group, the pions are systematically above the blast wave predictions for all energies. The $\phi$-mesons, which are supposed to behave differently from the other particles due to their smaller hadronic cross section, fit into the systematic of the other particles in that group.

In contrast to the behaviour seen in the particle group, the splitting of the data points among different anti-particle species decreases with increasing beam energy. At lower energies, the splitting for the anti-particle group is larger than for the particle group, but the difference between the two groups is decreasing with increasing beam energy. At $\sqrt{s_{\rm NN}} = $ 62.4 GeV the $v_{2}$ data for both groups, and accordingly the fits, are already very similar.

If we assume that the mass ordering in the low $p_{\rm T}$ region is only due to radial flow, then the difference in the splitting of particles and anti-particles indicates that the transverse expansion velocity is different for particles and anti-particles.
Figure~\ref{fig:beta} depicts the transverse expansion velocity $\beta$, which is extracted from the blast wave fits, as a function of beam energy with three different input temperatures as discussed in Sec.~\ref{sec_procedure}. The corresponding $\rho_{0}$ values are shown in Table~\ref{table}. The transverse expansion velocities for anti-particles are systematically higher than the ones for particles at all energies below $\sqrt{s_{\rm NN}} = $ 200 GeV, whereas the difference between particles and corresponding anti-particles decreases with increasing beam energy.
The latter is equivalent to the observation that the difference of $v_{2}$ between particles and anti-particles is decreasing with increasing beam energy~\cite{Adamczyk:2013}, therefore the transverse expansion velocity extracted from the blast wave fits becomes similar for both groups.

We also observe that the transverse expansion velocity for the particle group increases monotonically with energy, while the transverse expansion velocity for the anti-particle group decreases with energy up to $\sqrt{s_{\rm NN}} = $ 200 GeV, but then appears to increase, becoming identical with that of the particle group at $\sqrt{s_{\rm NN}} =$ 200 and 2760 GeV.

\subsection{Discussion}
\label{sec_discussion}

Qualitative explanations for a lower anti-particle $v_{2}$  compared to particles in the energy range of 7.7 $< \sqrt{s_{\rm NN}} <$ 39 GeV were discussed recently~\cite{MeanField,BayronStopping,Diffquark,isospin}. Various effects, like quark potentials or baryon stopping/baryon chemical potential, might be responsible for the observed difference in $v_{2}$. In the following we reconsider possible scenarios for different radial flow patterns for particles and anti-particles.

It is probable that anti-particle production at lower beam energies happens at the very early stage of the collision, where the energy density is high, either via thermal production or in a hard collision. Therefore, the produced anti-particles go through the whole expansion stage and get a larger transverse expansion velocity than the particles which could be produced at a later stage. At higher beam energies the production processes for particles and anti-particles becomes similar, which results in a smaller difference in $v_{2}(p_{\rm T})$ (Fig. 2 in Ref.~\cite{Adamczyk:2013}).

Figure~\ref{fig:netp} depicts the multiplicity rapidity density at mid-rapidity for 0\%--5\% central collisions for protons, anti-protons, and net-protons as a function of the center-of-mass energy $\sqrt{s_{\rm NN}}$. The anti-proton yield is monotonically rising with increasing $\sqrt{s_{\rm NN}}$, opposite to the net-proton yield ($dN/dy(p)$ - $dN/dy(\bar{p})$), which is decreasing. This is an indication for reduced baryon stopping at higher energies. The amount of stopped protons at $\sqrt{s_{\rm NN}} <$ 60 GeV exceeds the yield of produced protons. The proton $dN/dy|_{y_{\rm{cm}}}$ shows a minimum around that energy. It was speculated that the elliptic flow for produced and stopped particles might be different~\cite{BayronStopping}. A similar effect might be true for radial flow, which could explain the poorer description with the blast wave model of the particle group compared to the anti-particle group. The deviation between $\phi$-mesons and other particles might also be a result of the baryon stopping effect, which means $\phi$-mesons behave similar to other produced particles (anti-particles). In other words, the produced protons, which should follow the blast wave description, are contaminated at lower energies by stopped protons. In that case one should not trust to a certain level the results of a combined fit which includes particles with $u$ or $d$ quarks.

The few produced anti-protons in the collision center at lower energies might be annihilated due to the large absorption cross section and the large number of surrounding protons. Mainly anti-protons produced near the surface, where the radial flow is larger, may survive. This effect should decrease with increasing $\sqrt{s_{\rm NN}}$, where the $\bar{p}/p$ ratio is getting larger. Therefore the $\beta$ values for anti-particles are getting closer to the ones of particles. Figure~\ref{fig:beta} and Fig.~\ref{fig:netp} show that the anti-particle $\beta$ (0\%--80\%) is following the trend of the central proton $dN/dy$, which indicates a correlation between the two observables. The proton $\beta$ shows an opposite trend to net-proton. Because the transverse velocity extracted from particles are dominated by net-protons (stopped protons) at the lower beam energies, the net-protons show a smaller transverse velocity than produced protons. Proton and anti-proton transverse momenta spectra at the different energies and the use of finer collision centrality bins could shed light on the strength of the anti-proton absorption effect. 

\begin{figure}[h]
\centering
\resizebox{8cm}{!}{%
\includegraphics[scale=1]{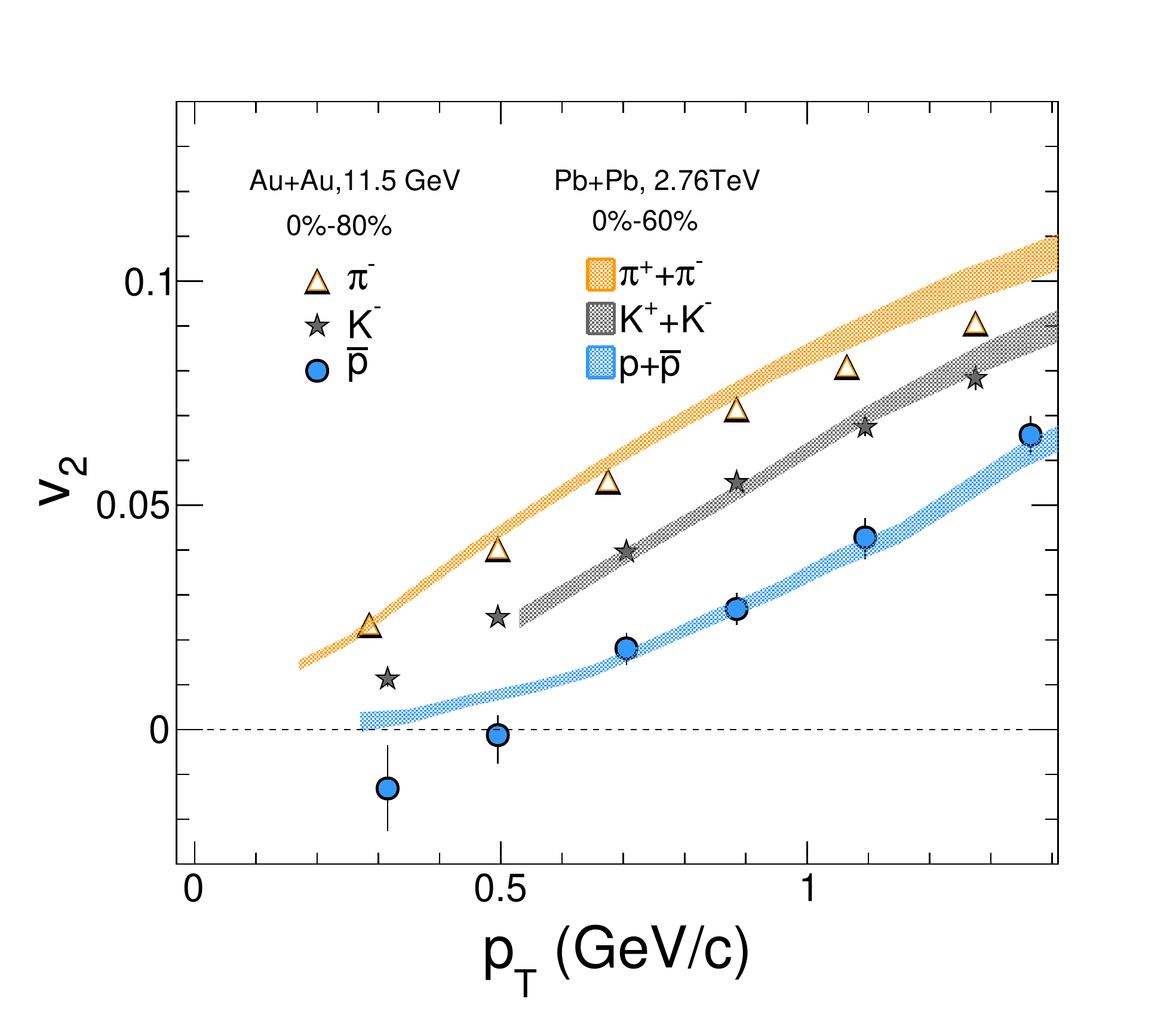}}
\caption{\label{fig:v2_comparison} (Color online) $v_{2}$ as a function of $p_{\rm T}$ from 0\%--80\% central Au+Au collisions at $\sqrt{s_{\rm NN}} = $ 11.5 GeV for the anti-particle group (points) and from 0\%--60\% central Pb+Pb collisions at $\sqrt{s_{\rm NN}} = $ 2.76 TeV (bands).} 
\end{figure}

The falling trend of $\beta$ for anti-particles from $\sqrt{s_{\rm NN}} = $ 7.7 to 200 GeV is opposite to the expectation for a hydrodynamic expanding system, which should show an increasing radial expansion velocity with increasing energy density. It is furthermore intriguing that the flow dependence on transverse momentum for anti-particles at $\sqrt{s_{\rm NN}} = $ 11.5 (0\%--80\%) and 2.76 TeV (0\%--60\%) are almost identical. The $v_{2}$ comparison can be found in Fig.~\ref{fig:v2_comparison}. In between, either $v_{2}$ at constant $p_{\rm T}$ is rising, or the boost in $p_{\rm T}$ for constant $v_{2}$ is decreasing with increasing energy. Both scenarios might be directly correlated since $v_{2}$ is an azimuthal modulation of the radial flow.

There is almost no difference of the $s_{2}$ and $\rho_{a}$ parameters between particles and anti-particles (shown in Table~\ref{table}). This is an indication that the driving force behind the difference of the $v_{2}$ values between particles and corresponding anti-particles in the low $p_{\rm T}$ region is due to the different $\beta$ parameters. On the other hand a consistent description of all anti-particles was achieved assuming that the radial expansion velocities for anti-protons and other anti-particles are different. In that case, other blast wave fit parameters than $\rho_{0}$ compensate this difference. A simultaneous fit of $v_{2}(p_{\rm T})$ and particle spectra would reduce those ambiguities. The scenario of different flow fields for particles and anti-particles shows the importance of a careful treatment of the initial and final state in future hybrid hydrodynamic calculations in the BES energy region. 
The $\rho_{a}$ parameter shows an increasing trend with increasing beam energy, which means the $v_{2}$ values should increase with increasing energy, like already observed in Ref.~\cite{Adamczyk:prc}. 

%
%
%
%
%
%
   

\section{Summary}
\label{sec_summary}
Simultaneous blast wave fits for $v_{2}(p_{\rm T})$, separated for particles and anti-particles, from 0\%--80\% (0\%--60\%) central Au+Au (Pb+Pb) collisions at $\sqrt{s_{\rm NN}} =$ 7.7--2760 GeV were presented. In general a reasonable description of the mass ordering of $v_{2}(p_{\rm T})$ in the low $p_{\rm T}$ range was achieved. We observed that blast wave fits for the anti-particle group are significantly better at $\sqrt{s_{\rm NN}} <$ 62.4 GeV compared to the particle group. Feed down effects were discussed, and it was shown that they might have a substantial impact on the observed small deviations form the blast wave expectation. The blast wave expectation for $\phi$ meson were shown to be systematically above the data for the particle group, whereas a consistent description for the anti-particles was attained. That might either show that $\phi$ mesons have a smaller radial or elliptic flow due to smaller hadronic cross section in comparison to the particle group, or that they follow the flow pattern of the anti-particle group which could be an indication for a distorted flow pattern in the particle group.
An energy dependent difference of the transverse expansion velocity $\beta$ between particles and corresponding anti-particles was observed at $\sqrt{s_{\rm NN}} <$ 62.4 GeV. $\beta$ is decreasing for anti-particles from $\sqrt{s_{\rm NN}}$ = 7.7--62.4 GeV, whereas the expansion velocity is monotonically increasing with $\sqrt{s_{\rm NN}}$ for the particle group. We discussed various effects, like anti-particle absorption, the early production of anti-particles, or the influence of stopped baryons on the radial flow of protons, which might explain the observed pattern. To distinguish those effects one needs the future particle spectra, finer centralities, and more statistics, especially for $\phi$ mesons at energies below $\sqrt{s_{\rm NN}}$ = 19.6 GeV. This will be achieved with the planned Beam Energy Scan II program at RHIC with a focus on energies below $\sqrt{s_{\rm NN}}$ = 20 GeV~\cite{BESII} and an expected increase in statistics of a factor 5--10.  

\section{Acknowledgements}
We thank Xin Dong, Ulrich Heinz, Volker Koch, Mike Lisa, Paul Sorensen, Sergei Voloshin and Nu Xu for important and useful discussions. This work was partly supported by the China Scholarship Council and by the Director, Office of Science, Office of Nuclear Science of the U.S. Department of Energy under Contract No. DE-AC02-05CH11231.
This work is also supported by the National Natural Science Foundation of China (NSFC U1332125) and the Program for Innovation Research of Science in Harbin Institute of Technology (PIRS OF HIT B201408).



\twocolumngrid


\end{document}